\documentclass[twocolumn,showpacs,floats,floatfix,superscriptaddress,aps,pra]{revtex4}

\usepackage{amsfonts,amssymb,amsmath}
\usepackage{color,calc}
\usepackage[dvips]{graphicx}
\usepackage{bm}

\def\be{\begin{equation}}
\def\ee{\end{equation}}
\def\bea{\begin{eqnarray}}
\def\eea{\end{eqnarray}}
\def\bse{\begin{subequations}}
\def\ese{\end{subequations}}

\def\ket#1{\vert #1 \rangle}
\def\bra#1{\langle #1 \vert}
\def\braket#1#2{\langle #1 \vert #2 \rangle}

\def\dicke{\ket{W^N_{N/2}}}
\def\1{\mathbf{1}}

\def\marked{s}
\def\j{j}

\begin{document}

\author{S. S. Ivanov}
\affiliation{Department of Physics, Sofia University, James Bourchier 5 blvd, 1164 Sofia, Bulgaria}
\affiliation{The School of Physics and Astronomy, University of Leeds, Leeds LS2 9JT, United Kingdom}

\author{P. A. Ivanov}
\affiliation{Department of Physics, Sofia University, James Bourchier 5 blvd, 1164 Sofia, Bulgaria}
\affiliation{Institut f\"{u}r Quanteninformationsverarbeitung, Universit\"{a}t Ulm, Albert-Einstein-Allee 11, 89081 Ulm, Germany}

\author{I. E. Linington\footnote{deceased}}
\affiliation{Department of Physics, Sofia University, James Bourchier 5 blvd, 1164 Sofia, Bulgaria}
\affiliation{Department of Physics and Astronomy, University of Sussex, Falmer, Brighton, BN1 9QH, United Kingdom}

\author{N. V. Vitanov}
\affiliation{Department of Physics, Sofia University, James Bourchier 5 blvd, 1164 Sofia, Bulgaria}
\affiliation{Institute of Solid State Physics, Bulgarian Academy of Sciences, Tsarigradsko chauss\'{e}e 72, 1784 Sofia, Bulgaria}

\title{Scalable quantum search using trapped ions}

\date{\today}

\begin{abstract}
We propose a scalable implementation of Grover's quantum search algorithm in a trapped-ion quantum information processor.
The system is initialized in an entangled Dicke state by using simple adiabatic techniques.
The inversion-about-average and the oracle operators take the form of \emph{single} off-resonant laser pulses,
addressing, respectively, all and half of the ions in the trap.
This is made possible by utilizing the physical symmetries of the trapped-ion linear crystal.
The physical realization of the algorithm represents a dramatic simplification:
each logical iteration (oracle and inversion about average) requires only \emph{two} physical interaction steps,
in contrast to the large number of concatenated gates required by previous approaches.
This does not only facilitate the implementation, but also increases the overall fidelity of the algorithm.
\end{abstract}

 \maketitle

\section{Introduction}

One of the most celebrated applications of quantum information processing is Grover's quantum search algorithm,
 which allows an initially unknown element to be determined from $\mathcal{N}$ equally likely possibilities
 in $O(\sqrt{\mathcal{N}})$ queries \cite{Grover}.
This outperforms the optimum classical strategy (a random 'trial and error' of elements),
 which requires $O\left(\mathcal{N}\right)$ steps on average.
In addition to providing a speed-up of the unstructured search problem,
 Grover's algorithm can also be adapted to look for solutions to a range of mathematically difficult problems that have structure,
  by nesting one quantum search inside another \cite{CGW}.
As with other applications of quantum computing, the benefits of a quantum over a classical approach increase with the size of the database.
Indeed, it has been suggested that the \emph{primary resource} for quantum computation is a Hilbert-space dimension,
 which grows exponentially with the available physical resources \cite{KCD}.
Another important consideration is that the physical and temporal resources required to implement the algorithm
 do not grow too rapidly with the register size $\mathcal{N}$.

Proof-of-principle quantum search has been successfully demonstrated in nuclear magnetic resonance \cite{CGK,JMH,MSA},
 linear optical \cite{K,Walther} and trapped-ion systems \cite{Brickman},
 as well as with individual Rydberg atoms \cite{AWB} and in classical optics \cite{Bhattacharya}.
Of these, only the trapped-ion platform possesses a fully scalable Hilbert space
 and in this sense it is realistically the only candidate for performing a practically useful quantum search.
We note, however, that while the trapped-ion system is scalable \cite{Kielpinski},
 the largest dimensional quantum search so far performed with trapped ions was for a database size of $\mathcal{N}=4$ \cite{Brickman}.
Extending the approach of Ref. \cite{Brickman} to a large number of ions is highly demanding,
 since it requires the ability to construct, with very high fidelity, a great number of multiply-conditional gates.
For two ions, as in Ref. \cite{Brickman}, each oracle query amounts to a controlled phase gate between the ions.
When $N$ ions are involved, however, the oracle operator becomes a gate which is multiply-conditional upon the internal state of \emph{all} $N$ \emph{ions}.
Although such a multiply-conditional gate may be decomposed efficiently using a series of one- and two-ion gates \cite{Barenco},
 in practice such a synthesis becomes a daunting task, even for a moderate number of ions.
For example, for a four-qubit register, the oracle may be constructed using 13 two-qubit conditional gates \cite{Barenco},
each of these requiring 5 consecutive physical interactions on average,
 which is beyond the capabilities of current experiments \cite{CNOT,CP,TG}.
The above example makes it apparent that there is a clear distinction between an approach,
 which is only \emph{formally scalable},
 and an approach, which is \emph{realistically scalable} using current technology.

It is highly desirable, therefore, to find new ways of performing quantum search in a scalable system,
 which does not require an exorbitant number of elementary physical interactions.
In this paper we propose a novel approach
 to perform Grover's search algorithm.
The linear crystal of $N$ ions is prepared in a symmetric Dicke state with $N/2$ excitations, $\dicke$;
 hence the register dimension scales \emph{exponentially} with the number of ions.
The implementation of the quantum search involves a series of red-sideband laser pulses
 addressing all, or half, of the ions in the trap.
Thus each of the inversion about average and the oracle query can be produced in a \emph{single} interaction step.
Consequently, the total number of \emph{physical} steps is the same as the number of algorithmic steps.

The implementation of the Grover's search algorithm proposed below follows earlier proposals,
wherein the database scales linearly \cite{GroverLin} (see also \cite{Kyoseva09})
and quadratically \cite{GroverQuad} with the number of ions $N$.
The database size here, which is $\mathcal{N}\sim 2^N \sqrt{2/(\pi N)}$, scales exponentially with $N$ and it is therefore of fully quantum nature.
The reduction of the full Hilbert space of dimension $2^N$ by a factor of $\sqrt{N}$
is compensated by the absence of any ancilla qubits or the need of error correction
because of the ultra-compact implementation, in which each logical mathematical step is realized with a single physical interaction (laser pulse).

The remainder of this paper is organized as follows.
A brief review of Grover's algorithm for a quantum mechanical
speed-up of the unstructured search problem is given in Sec. \ref{QuantumSearch}.
Particular attention is given to the ideas of amplitude amplification and generalized reflections,
which lie at the heart of the search algorithm.
Sec. \ref{System} introduces a model Hamiltonian describing a single laser pulse illuminating a chain of trapped ions,
and a convenient Hilbert-space factorization, with which to describe the dynamics.
By making use of the simplified dynamics in this factorized basis, we construct
recipes for the approximate synthesis of the two reflections required for a quantum search,
and examples of numerical simulations are presented in Sec. \ref{Algorithm}.
In Sec. \ref{Conclusions} we summarize our findings.

\section{Grover's search algorithm}\label{QuantumSearch}

Grover's algorithm provides a method for solving the unstructured search problem, which can be stated as follows: given a collection of database
elements $x=1,2,\ldots,\mathcal{N}$, and an \emph{oracle function} $f(x)$ that acts differently on one \emph{marked} element $\marked$ to all others,
\begin{equation}
f(x) =\left\{\begin{array}{ll}
1,& x=\marked, \\
0,& x\neq \marked,
\end{array}\right.  \label{oracle_fun}
\end{equation}
find the marked element in as few calls to $f(x)$ as possible \cite{Grover}.
If the database is encoded in a physical system that behaves classically, then each oracle query can only act on a single database element.
In this case, the optimal search strategy is simply a random selection of elements; on average, it will be necessary to make
approximately $\mathcal{N}/2$ calls to the oracle before the marked element $\marked$ is located.
The idea underlying Grover's algorithm is to encode the database in a physical system that behaves
quantum mechanically. Therefore, each possible search outcome is represented as a basis vector
$\ket{x}$ in an $\mathcal{N}$-dimensional Hilbert space;
correspondingly, the marked element is encoded by a \emph{marked state} $\ket{\marked}$. Hence one can apply unitary operations
(involving the oracle function) to \emph{superpositions} of the different search outcomes.
It is thus possible to amplify the amplitude of the marked state $\ket{\marked}$
using constructive interference, while attenuating all other amplitudes, and locate the marked element in $O(\sqrt{\mathcal{N}})$ steps.
Before the execution of the algorithm, the quantum register is prepared in an \emph{equal} superposition of all basis elements,
\begin{equation}
\left\vert W\right\rangle =\frac{1}{\sqrt{\mathcal{N}}}\sum_{x=1}^{\mathcal{N%
}}\left\vert x\right\rangle .  \label{W_register}
\end{equation}

Central to the operation of the quantum search algorithm is the idea of generalized complex reflections,
known in the computer science literature as \emph{Householder reflections} (HR) \cite{ASH}:
\begin{equation}
\hat{M}_{\psi } (\phi) = \mathbf{1}+ (e^{i\phi}-1) \ket{\psi} \bra{\psi} .  \label{GHR}
\end{equation}
When the phase $\phi$ is set equal to $\pi$, the effect of $\hat{M} _{\psi}(\phi)$ on any vector
is to invert the sign of the component of this vector along $\ket{\psi}$, while leaving all other components unchanged,
which amounts to a reflection with respect to an $(\mathcal{N-}1)$-dimensional plane orthogonal to $\ket{\psi}$.
Allowing $\phi$ to take arbitrary values extends the concept of reflection, 
by imprinting an arbitrary phase onto the component along $\ket{\psi}$, rather than a simple sign inversion.
Householder reflections are widely used in classical data analysis
 and also constitute a powerful tool for coherent manipulation of quantum systems \cite{IKV,ITV,IV,Kyoseva,drando}.
The core component of Grover's algorithm is a pair of coupled Householder reflections, which together form a single \emph{Grover operator} $\hat{G}$:
\begin{equation}
\hat{G}=\hat{M}_{W}\left(\varphi_W \right) \hat{M}_{\marked} (\varphi_{\marked}). \label{Grover_op}
\end{equation}
According to standard nomenclature, the operator $\hat{M}_{\marked} (\varphi_{\marked})$ is referred to as the oracle query,
while $\hat{M}_{W}(\varphi_W)$ is known as the inversion-about-average operator.

We note that with the initial state given in Eq.~\eqref{W_register}, and during successive applications of the operator $\hat{G}$,
the state vector for the system begins and remains in the two-dimensional subspace defined by the non-orthogonal states $\ket{\marked}$ and $\ket{W}$. Each application of $\hat{G}$ amplifies the marked state population until it reaches a maximum value close to unity after $n_{G}$ iterations,
at which point the search result can be read out by a measurement in the computational basis.

The problem of how to optimize the quantum search routine by allowing arbitrary $\varphi_W$ and $\varphi_\marked$
has been studied extensively \cite{Long1,ketaici,Hoyer}.
It is found that the maximum possible amplitude amplification per step of the marked state arises
when the phases $\varphi_W$ and $\varphi_\marked$ are both set to $\pi$ (as in Grover's original proposal \cite{Grover}).
The corresponding minimum number of search steps $n_{G}^{\min}$ is given by:
\begin{equation}\label{n_min}
n_{G}^{\min} = \left[\frac{\pi}{2\arcsin\left(2\sqrt{\mathcal{N}-1}/\mathcal{N}\right)}\right]
 \overset{\mathcal{N}\gg 1}{\sim} \left[\frac{\pi\sqrt{\mathcal{N}}}{4}\right],
\end{equation}
where $[n]$ denotes the integer part of $n$.
However, this choice of phases is not unique.
For large $\mathcal{N}$, as long as the \emph{phase matching} condition $\varphi_W=\varphi_\marked=\varphi$ is satisfied \cite{ketaici},
a high fidelity search can be performed for any value of $\varphi$ in the range $0<\varphi\leqq\pi$
and for certain values of $\varphi$, a deterministic quantum search is possible \cite{Long1}.

In the following two sections, we shall describe how to perform a
physically efficient quantum search using the dynamic symmetries in a system of trapped ions.

\section{Ion-Trap Implementation}\label{System}

\subsection{Model Hamiltonian}

We consider $N$ ions confined and laser cooled in a linear Paul trap, each with two relevant internal states $\ket{0}$ and $\ket{1}$,
 with respective transition frequency $\omega_{0}$.
The linear ion crystal 
 interacts with a laser pulse with frequency $\omega_{\textrm{L}}=\omega_{0}-\omega_{\textrm{tr}}-\delta$,
 where $\omega_{\textrm{tr}}$ is the axial trap frequency and $\delta$ is the laser detuning from the first red-sideband resonance.
We assume that the phonon spectrum can be resolved sufficiently well that only the center-of-mass mode is excited by
 this interaction and that other vibrational modes can safely be neglected.
In the Lamb-Dicke limit, and after making the optical and vibrational rotating-wave approximations, the interaction Hamiltonian is \cite{Wineland}:
\begin{equation}
\hat{H}_{I}(t) = \hbar\sum_{k=1}^{N}\left[ g\left( t\right) \left(\sigma _{k}^{+}\hat{a}+\sigma _{k}^{-}\hat{a}^{\dag }\right)
+\frac{\delta }{2}\sigma _{k}^{\left( z\right) }\right] .
\label{Hamiltonian}
\end{equation}
Here $\sigma _{k}^{+}=|1_{k}\rangle\langle0_{k}|$ and $\sigma_{k}^{-}=|0_{k}\rangle\langle1_{k}|$ are the raising and lowering
 operators for the internal states of the $k$th ion, $\sigma_{k}^{\left( z\right)}$ is the Pauli spin matrix
 and $\hat{a}^{\dag}$ and $\hat{a}$ are respectively the creation and annihilation operators of center-of-mass phonons.
The coupling between the internal and motional degrees of freedom is $g(t) =\eta \Omega(t)/2\sqrt{N}$,
 where $\eta =\sqrt{\hbar |\textbf{k}|^{2}/2M\omega _{\textrm{tr}}}$ is the single-ion Lamb-Dicke parameter,
 with $\textbf{k}$ being the laser wave vector and $M$ is the mass of the ion.
The function $\Omega(t)$ is the real-valued time-dependent Rabi frequency.

The Hamiltonian \eqref{Hamiltonian} conserves the total number of excitations ($n_i$ ionic plus $n_p$ motional), which in the scheme we propose, is
half the number of ions, i.e. $N/2$ (with $N$ even), $n_i+n_p=N/2$. The energy pattern splits into manifolds corresponding to $n_{i}$ ionic and $n_{p}=N/2-n_i$ motional excitations.
Each manifold is $C_{n_{i}}^{N}$-fold degenerate, where $C_{n_{i}}^{N}=N!/n_{i} !\left( N-n_{i}\right)!$.
It is readily verified that the dimension of the $n_i=N/2$ manifold, $\mathcal{D}$,
grows \emph{exponentially} with $N$; indeed, for large $N$ we have
\be\label{database size}
\mathcal{N}\equiv C_{N/2}^{N}\sim \frac{2^N}{\sqrt{\pi N/2}}
\left[1 - \frac{1}{4N} + \mathcal{O}\left(N^{-2}\right)\right].
\ee
The subspace of the overall Hilbert space, which spans the manifold $\mathcal{D}$, we shall use to represent the state of the register in Sec. \ref{Algorithm}; $\mathcal{D}$ is the set of states, which encode the database of $\mathcal{N}$ elements.

\subsection{Hilbert space factorization}

\begin{figure}[tb]
\includegraphics[angle=0,width=0.9\columnwidth]{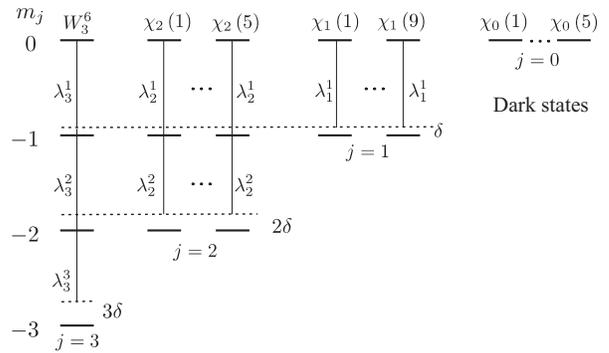}
\caption{
The MS basis states for $N=6$ ions with 3 excitations consists of a series of independent chainwise linkages.
Since the total number of excitations is half the number of ions, then from Eq. \eqref{mj} follows that states with $m_j>0$
are inaccessible and are therefore not shown.
The states that make up the longest ladder are all symmetric Dicke states \cite{LV1}.
The number of motional $n_{p}$ and ionic $n_{i}$ excitations for each level can be inferred from $m_j$, Eq. \eqref{mj}.}
\label{fig1}
\end{figure}

It is possible to move to a new basis in
which the Hilbert space is factorized into a collection of noninteracting chains of states.
The new states we shall call MS states, since they can be obtained by the multilevel Morris-Shore (MS) factorization \cite{RVS}.

To determine the MS states first we rewrite the Hamiltonian \eqref{Hamiltonian} in terms of the total ionic pseudospin:
\begin{equation}
\hat{H}_{I}(t)=\hbar g\left( t\right) \left(
\hat{a}\hat{J}_{+}+\hat{a}^{\dag }\hat{J}_{-}\right) +\hbar\delta
\hat{J}_{z}, \label{Hamiltonian1}
\end{equation}
where $\hat{J}_{\pm} =\sum_{k=1}^{N}\sigma _{k}^{\pm}$
and $\hat{J}_{z}=\frac{1}{2}\sum_{k=1}^{N}\sigma _{k}^{\left(
z\right)}$.
The MS basis consists of the set of the eigenvectors of the two
commuting operators $\hat{J}^2$ and $\hat{J}_{z}$, where $\hat{J}^2=\frac{1}{2}(\hat{J}_{+}\hat{J}_{-}+\hat{J}_{-}\hat{J}_{+})+\hat{J}_z^2$.
Therefore each MS state is assigned two quantum numbers,
respectively $j$ and $m_j$. Since $\hat{J}^2$ commutes also with the Hamiltonian \eqref{Hamiltonian1},
the Hilbert space factorizes into a set of decoupled chains with different values of $j$.

The MS states are sketched in Fig. \ref{fig1} for 6 ions with 3 excitations.
Because the laser pulse couples equally to every ion in the trap,
the longest chain is comprised of the symmetric Dicke states $\ket{W^N_{n_i}}$,
wherein a given number of ionic excitations $n_i$ is shared evenly amongst all the ions
in the trap \cite{LV1}. Each chain is assigned different $j$ and is comprised of
states with different $m_j$, which varies from 0 to $j$, corresponding to the number of
excited ions. If $n_{i}$ of the ions are excited, then 
\begin{equation}
m_{j}=n_{i} -\frac{N}{2}=-n_p.  \label{mj}
\end{equation}
Since the total number of excitations is half the number of ions, then from Eq. \eqref{mj} it follows that for each chain
$m_j$ has a maximum value of 0, i.e. states with $m_j>0$ are inaccessible.
By analogy with the traditional angular momentum operators, it follows that the number of states in a chain is equal to $j+1$ ($j$ of the states are
not accessible), from which $j$ can be inferred.
The longest chain corresponds to $j=N/2$, the next longest to $j=N/2-1$ and so on,
and overall, there are $\left[N/2+1\right]$ chains of different length in the factorized coupling scheme.

For the following analysis, it will be necessary to
go further and calculate the couplings in the MS basis.
The coupling between the
neighbors $\ket{j,m_j}$ and $\ket{j,m_j-1}$ follows immediately
from the matrix elements of the operators $\hat{a}^{\dag}$ and
$\hat{J}_{-}$, i.e.
\begin{align}
\lambda _{\j }^{m_{j}} (t) &= g(t) \sqrt{n_{p} (j+m_{j}) (j-m_{j}+1)}  \notag \\
&= g(t) \sqrt{n_{p}(j+n_i-N/2) (j+N/2+1-n_i)}\,.  \label{coupling}
\end{align}

As illustrated in Fig. \ref{fig1}, there can be many different degenerate MS states with the same values of $\j$ and $m_{j}$.
We label the states with a given $\j\neq N/2$ and $m_j=0$ with $\ket{\chi_{\j}(k)}$,
where $k=1,\ldots,N_{\j}$, with $N_{\j}=C_{N/2-j }^{N}-C_{N/2-j -1}^{N}$.

\section{Implementation of Grover's algorithm}\label{Algorithm}

The manifold $\mathcal{D}$, which encodes the database elements, consists of the states,
for which the pseudo-angular momentum projection is $m_{j}=0$, Eq. \eqref{mj}.
This requires that the total number of ions $N$ is even.
Half of these are in state $\ket{1}$, while the other half are in state $\ket{0}$ ($n_{i}=N/2$).
The number of elements $\mathcal{N}$ in the database therefore scales exponentially with the number of ions $N$, Eq.~\eqref{database size}.
The elements $\ket{x}$ in Eq.~\eqref{W_register}, which belong to $\mathcal{D}$, can now be written as
\begin{equation}
\ket{x} = P_{x}\ket{ 1_{1}\ldots 1_{N/2}0_{N/2+1} \ldots 0_{N}} ,  \label{x}
\end{equation}
where the subscript $x$ runs over all distinct permutations $P_{x}$ of the ions' internal states.
The initial state $\ket{W}$ is thus a symmetric Dicke state $\dicke$ of $N$ ions sharing $N/2$ excitations,
\begin{equation}
\dicke = \frac{1}{\sqrt{C_{N/2}^{N}}}\sum_{x}P_{x}\left\vert 1_{1}\ldots 1_{N/2 }0_{N/2+1}\ldots 0_{N}\right\rangle .
\label{Dicke}
\end{equation}
Our proposed experimental procedure consists of four simple operations.
(i) The ions are firstly initialized in the entangled Dicke state \eqref{Dicke}.
This may be achieved using very simple adiabatic passage techniques, involving either a pair
 of chirped laser pulses \cite{LV1,Hume} or two pairs of delayed but overlapping laser pulses \cite{LV2}, and using global addressing.
(ii) Synthesis of the inversion-about-average operator is appealingly simple:
 $\hat{M}_{W}$ is a single red-sideband off-resonant laser pulse
addressing all ions in the linear chain.
(iii) The oracle query $\hat{M}_{\marked}$ is also a single red-sideband laser pulse, applied on half of the ions.
After an appropriate number of iterations \eqref{n_min}, the system evolves into the marked state $\ket{\marked}$,
which can be identified by performing (iv) a fluorescence measurement on the entire chain.

\subsection{Synthesis of the inversion-about-average operator}

In most existing proposals for implementing Grover's search algorithm using trapped ions the generation of $\hat{M}_{W}$
 requires a large number of concatenated physical interaction steps, even for moderate register size $\mathcal{N}$.
However, by restricting the dynamics to the subspace of the overall Hilbert space
 in which only half of the ions are excited, this operator becomes possible to synthesize in only a \emph{single} interaction step.
This simplification is achieved by taking advantage of the fact that both the Hamiltonian \eqref{Hamiltonian1}
 and the state $\dicke$ are symmetric under exchange of any two ions.

The energies of the MS states do not cross in time so that in the limit $\delta \gg 1/T$
 the transitions between the MS states vanish due to the effect of adiabatic complete population return \cite{Vitanov}.
Each of the MS states acquires a phase shift $\varphi_{\j}$ (the index $j$ corresponds to the eigenvalue of $\hat{J}^2$) and the unitary propagator within the Dicke manifold $\mathcal{D}$
 is a product of $C_{N/2-1}^{N}$ coupled reflections \cite{Kyoseva}
\begin{equation}
\hat{U}_{W} = \hat{M}_{W}(\varphi_{W}) \prod\limits_{j=1}^{N/2 -1}
 \prod\limits_{k=1}^{N_{\j}} \hat{M}_{\chi_{\j}(k)}(\varphi _{\j}), \label{propagator}
\end{equation}
with $N_j=C^{N}_{N/2-j}-C^{N}_{N/2-j-1}$ and $\varphi_W=\varphi_{N/2}$.
For a given value of the coupling, the detuning may be adjusted in order to control the phases $\varphi_{\j}$.
Ideally, these interaction parameters would be chosen such that $\varphi_{W}\neq 0$ (e.g., $\varphi_{W}=\pi$),
while $\varphi_{j\neq N/2}=0$ as this would result in $\hat{U}_{W}$ being identical to the inversion-about-average operator $\hat{M}_{W}(\varphi_{W})$.

\subsection{Synthesis of the oracle operator}

The effect of each oracle query $\hat{M}_{\marked}(\phi_{\marked})$ is to imprint a phase shift $\phi_{\marked}$ on the marked state $\ket{\marked}$,
 whilst leaving all other computational basis states unchanged.
In general this can be achieved by a multiply-conditional phase gate upon the internal state of all $N$ ions in the trap.
When more than a few ions are involved, this becomes a prohibitively complicated operation,
which generally requires an immense number of one- and two-qubit gates \cite{Barenco}.
However, since we do not work in the whole Hilbert space,
but rather in the manifold $\mathcal{D}$,
the oracle operator can be implemented in a much simpler fashion -- by a \emph{single} red-sideband laser pulse,
addressing uniformly those $N/2$ ions in the trap, which share the excitation of the marked state.

Let's consider an example, when the marked state is $\ket{\marked}=\ket{111000}$. Then we address the first three ions.
Since the initial state $\dicke$ and the interactions with the laser are symmetric under exchange of the first three ions, and the last three, as well,
during the execution of the algorithm the state of the system is a linear combination of the states $\ket{\Phi_{k}}=\ket{W_{N/2-k}^{N/2}}\ket{W_{k}^{N/2}}$,
$k=0,\ldots,N/2$; the marked state is $\ket{\marked}=\ket{\Phi_0}$. Because the Hamiltonian, which describes the oracle call,
does not change the total number of excitations of the first (and the last) three ions, i.e. it does not drive the system between $\ket{\Phi_{k_1}}$ and $\ket{\Phi_{k_2}}$ for $k_1\neq k_2$,
in the adiabatic limit the states $\ket{\Phi_k}$ acquire only phase shifts $\phi_k$. Therefore the propagator in the manifold $\mathcal{D}$ is formally described by the action of $N/2$ coupled reflections,
\begin{equation}\label{propagator_oracle}
\hat{U}_{\marked} = \hat{M}_{\marked}(\phi_{\marked}) \prod\limits_{k=1}^{N/2-1}\hat{M}_{\Phi_{k}}(\phi_{k}).
\end{equation}
Hence, as in the case of the inversion-about-average operator, we need to control $N/2$ phases.
The interaction parameters should be chosen such that $\phi_{\marked}=\varphi_{W}$, while $\phi_{k\neq N/2}=0$,
 as this would result in $\hat{U}_{\marked}$ being identical to the oracle operator $\hat{M}_{\marked}(\phi_{\marked})$.

\subsection{Phase conditions}

The phase conditions for the propagators $\hat{U}_W$ and $\hat{U}_s$, derived above, cannot be satisfied exactly, since the
phases $\varphi_k$ and $\phi_k$ are overly commensurate.
However one can still perform the algorithm with sufficiently high fidelity
\begin{equation}
P=1-2\left|\text{Re}\braket{f}{\Delta f}\right|.
\label{infidelity}
\end{equation}
Here $\ket{f}$ is the state of the system after the completion
of the algorithm and $\ket{\Delta f}$ is its deviation due to the phases deviation.
Fortunately, a Fourier expansion reveals that phase deviation of all odd orders
leads to the purely imaginary bracket $\braket{f}{\Delta f}$ and hence do not contribute to Eq.~\eqref{infidelity}.
As a result, the algorithm is less sensitive to phase deviations
since only those of second order have a leading affect on the final state populations.

\begin{figure}[tb]
\includegraphics[angle=270,width=0.9\columnwidth]{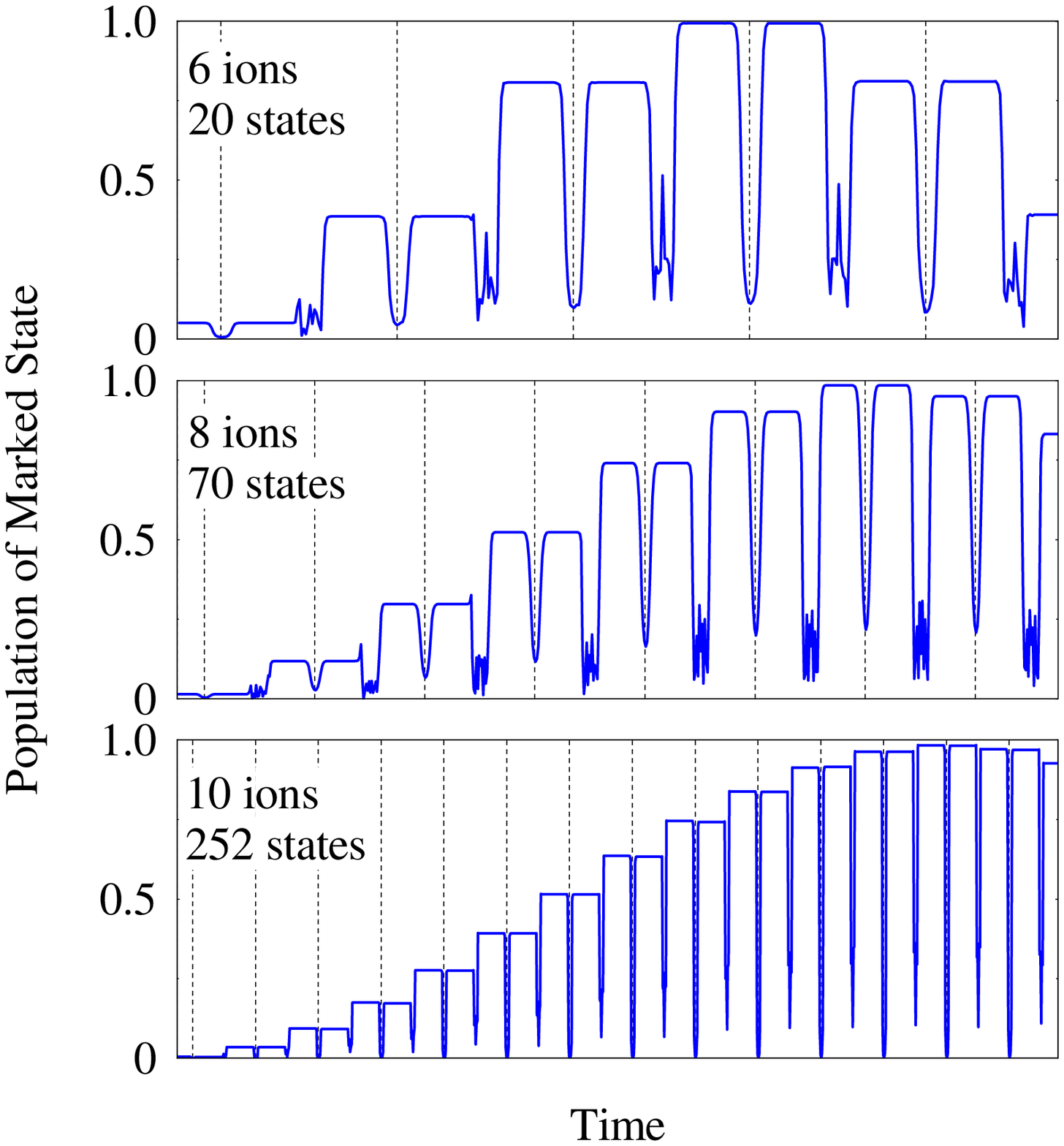}
\caption{(Color online) Simulation of the Grover search algorithm with $N$ ions and $N/2$ excitations (register dimension $C^N_{N/2}$),
for $N=6$ (upper frame), $N=8$ (middle frame) and $N=10$ (lower frame).
The system of ions is assumed to be initialized in the symmetric Dicke state $\ket{W^{N}_{N/2}}$.
The laser pulses have a Gaussian shape, $g(t) = g_0 e^{-t^2/T^2}$.
The thin vertical lines display the timing of the oracle call.
The computed numerically scaled detunings and peak Rabi frequencies for the oracle and the inversion-about-average operator,
respectively, are given in the table below.
The marked-state population (the fidelity) of around $99\%$ is obtained after $n_g=3$, 6 and 12 steps,
respectively, in exact agreement with Grover's value \eqref{n_min}. The
numerical simulation includes all off-resonant transitions to states with $m_j\neq 0$.}
\begin{center}\begin{tabular}{|r|r|r|cc|cc|}\hline
\#ions & \#elements & \#steps & oracle & & reflection & \\
$N$ & $\mathcal{N}$ & $n_G$ & $\delta T$ & $g_0 T$ & $\delta T$ & $g_0 T$ \\ \hline
6 & 20 & 3 & 19.470 & 28.610 & 10.320 & 25.830 \\
8 & 70 & 6 & 21.400 & 10.800 & 21.050 & 24.400 \\
10 & 252 & 12 & 88.565 & 87.142 & 15.687 & 70.322 \\
\hline
\end{tabular}\end{center}
%
\label{fig2}
\end{figure}

\subsection{Numerical Demonstration}

As a demonstration, we have solved the Schr\"{o}dinger equation numerically for a Gaussian pulse shape
and a constant detuning $\delta$.
Sample results are shown in Fig. \ref{fig2} for $N=6$, 8 and 10 ions with 3, 4 and 5 excitations,
 respectively, which imply databases of $\mathcal{N}=20$, 70 and 252 elements.
The fidelity plotted on the vertical axis is the time-dependent population of the marked state.
The system of ions is assumed to be prepared initially in the Dicke superposition $\dicke$ of ionic collective states, each of which contains
 exactly $N/2$ ion qubits in state $\ket{1}$ and $N/2$ in state $\ket{0}$.
Each Grover iteration consists of a phase shift for the marked state (oracle call), which amounts a single red-sideband laser pulse,
 addressing uniformly those $N/2$ ions in the chain, which share the excitation of the marked state, followed by a single red-sideband pulse,
 which addresses uniformly the entire ion chain (the inversion-about-average operator).
The number of steps, for which the algorithm singles out the marked item with a probability of around $99\%$,
 is respectively $n_g=3$, 6 and 12 as predicted by Eq.~\eqref{n_min}.

\section{Conclusions}\label{Conclusions}

Despite the intense flurry of theoretical and experimental activity in the
decade following Grover's original proposal for a quantum speed-up of
unstructured search, a large discrepancy still remains between the current
experimental state-of-the-art and what is required for a practically useful quantum search.
To highlight this incongruence, we note that the only physical system in which a scalable quantum search has been performed is in
 a chain of trapped ions, and in that experiment, the search-space totalled just four elements.
An important intermediate goal on the long road to
performing a practically useful quantum search is to demonstrate Grover's algorithm in a moderately-sized trapped-ion quantum register.
The standard approach of building up the many-ion Grover operator using a network of single- and two-qubit gates is inappropriate for this task,
 since the required physical resources far exceed those available in today's experiments.
By contrast, in this article, we have proposed to construct
the many-ion Grover operator using only two single off-resonant laser pulses, with a suitably chosen peak Rabi frequencies and detunings,
which synthesize the inversion-about-average and oracle operators, each in a single shot.

A simple recipe for synthesizing the inversion-about-average operator was
derived by factorizing the overall Hilbert space into a series of
independent ladders.
The coupling strengths between the MS states were determined solely through a
consideration of the angular-momentum structure of the combined ionic
pseudospin.
The technique proposed in this paper raises the
prospect of demonstrating Grover's algorithm in a moderately sized trapped-ion
database comprising up to several hundred elements, and which scales
exponentially with the number of ions; this is a necessary step on the path
to demonstrating a practically useful quantum search, which remains a long-term goal.

This work is supported by European Commission's projects EMALI and FASTQUAST, the Bulgarian Science Fund grants VU-F-205/06,
D002-90/08, Sofia University grant 020/2009, and Elite programme of the Landesstiftung-Baden-W\"{u}rttemberg.



\begin{thebibliography}{99}
\bibitem{Grover} L. K. Grover, Phys. Rev. Lett. \textbf{79}, 325 (1997).

\bibitem{CGW} N. J. Cerf, L. K. Grover, and C. P. Williams, Phys. Rev. A \textbf{61}, 032303 (2000).

\bibitem{KCD} R. Blume-Kohout, C. M. Caves, and I. H. Deutsch, Found. Phys. \textbf{32} (2002).

\bibitem{CGK} I. L. Chuang, N. Gershenfeld, and M. Kubinec, Phys. Rev. Lett. \textbf{80}, 3408 (1998).

\bibitem{JMH} J. A. Jones, M. Mosca, and R. H. Hansen, Nature \textbf{393}, 344 (1998).

\bibitem{MSA} M. S. Anwar, D. Blazina, H. Carteret, S. B. Duckett, J. A. Jones, Chem. Phys. Lett. \textbf{400}, 94 (2004).

\bibitem{K} P. G. Kwiat, J. R. Mitchell, P. D. D. Schwindt, A. G. White, J. Mod. Opt. \textbf{47}, 257 (2000).

\bibitem{Walther} P. Walther, K. J. Resch, T. Rudolph, E. Schenck, H. Weinfurter, V. Vedral, M. Aspelmeyer, and A. Zeilinger, Nature \textbf{434}, 169 (2005).

\bibitem{Brickman} K.-A. Brickman, P. C. Haljan, P. J. Lee, M. Acton, L. Deslauriers, and C. Monroe, Phys. Rev. A \textbf{72}, 050306(R) (2005).

\bibitem{AWB} J. Ahn, T. C. Weinacht, and P. H. Bucksbaum, Science \textbf{287}, 463 (2000).

\bibitem{Bhattacharya} N. Bhattacharya, H. B. van Linden van den Heuvell, and R. J. C. Spreeuw, Phys. Rev. Lett. \textbf{88}, 137901 (2002).

\bibitem{Kielpinski} D. Kielpinski, C. Monroe, D. J. Wineland, Nature \textbf{417}, 709 (2002).

\bibitem{Barenco} A. Barenco, C. H. Bennett, R. Cleve, D. P. DiVincenzo, N. Margolus, P. Shor, T. Sleator, J. A. Smolin, and H. Weinfurter, Phys. Rev. A \textbf{52}, 3457 (1995).

\bibitem{CNOT} F. Schmidt-Kaler, H. H\"{a}fner, M. Riebe, S. Gulde, G. P. T. Lancaster, T. Deuschle, C. Becher, C. F. Roos, J. Eschner and R. Blatt, Nature \textbf{422}, 408 (2003).

\bibitem{CP} D. Leibfried, B. DeMarco, V. Meyer, D. Lucas, M. Barrett, J. Britton, W. M. Itano, B. Jelenkovic, C. Langer, T. Rosenband and D. J. Wineland, Nature \textbf{422}, 412 (2003).

\bibitem{TG} T. Monz, K. Kim, W. Hansel, M. Riebe, A. S. Villar, P. Schindler, M. Chwalla, M. Hennrich, and R. Blatt, Phys. Rev. Lett. \textbf{102}, 040501 (2009).

\bibitem{Haffner} H. H\"{a}ffner, C. F. Roos and R. Blatt, Phys. Rep. \textbf{469}, 155 (2008).

\bibitem{GroverLin} S. S. Ivanov, P. A. Ivanov, and N. V. Vitanov, Phys. Rev. A \textbf{78}, 030301(R) (2008).

\bibitem{GroverQuad} I. E. Linington, P. A. Ivanov, and N. V. Vitanov, Phys. Rev. A \textbf{79}, 012322 (2009).

\bibitem{Kyoseva09} E. S. Kyoseva, D. G. Angelakis, L.C. Kwek, arXiv:0908.3308v1 [quant-ph], to appear in Europhys. Lett.

\bibitem{ASH} A. S. Householder, J. ACM \textbf{5}, 339 (1958).

\bibitem{IKV} P. A. Ivanov, E. S. Kyoseva, and N. V. Vitanov, Pys. Rev. A \textbf{74}, 022323 (2006).

\bibitem{ITV} P. A. Ivanov, B. T. Torosov, and N. V. Vitanov, Pys. Rev. A \textbf{75}, 012323 (2007).

\bibitem{IV} P. A. Ivanov and N. V. Vitanov, Pys. Rev. A \textbf{77}, 012335 (2008).

\bibitem{Kyoseva} E. S. Kyoseva,  N. V. Vitanov, B. W. Shore, J. Mod. Opt \textbf{54}, 2237 (2007).

\bibitem{drando} A. A. Rangelov, N. V. Vitanov, and B. W. Shore, Phys. Rev. A \textbf{77}, 033404 (2008).


\bibitem{Long1} G. L. Long, Phys. Rev. A \textbf{64}, 022307 (2001).

\bibitem{ketaici} G. L. Long, Y. S. Li, W. L. Zhang, L. Niu, Phys. Lett. A \textbf{262} (1999) 27-34.

\bibitem{Hoyer} P. Hoyer, Phys. Rev. A \textbf{62}, 052304 (2000).

\bibitem{Wineland} D. J. Wineland, C. Monroe, W. M. Itano, D. Leibfried, B. E. King, and D. M. Meekhof, J. Res. Natl. Inst. Stand. Technol. \textbf{103}, 259 (1998).

\bibitem{LV1} I. E. Linington and N. V. Vitanov, Phys. Rev. A \textbf{77}, 010302(R) (2008).

\bibitem{LV2} I. E. Linington and N. V. Vitanov, Phys. Rev. A \textbf{77}, 062327 (2008).

\bibitem{Retzker} A. Retzker, E. Solano, and B. Reznik, Phys. Rev. A \textbf{75}, 022312 (2007).

\bibitem{RVS} A. A. Rangelov, N. V. Vitanov, and B. W. Shore, Phys. Rev. A \textbf{74}, 053402 (2006).

\bibitem{MW} L. Mandel and E. Wolf, \emph{Optical Coherence and Quantum Optics} (Cambridge University Press, 1995).

\bibitem{Hume} D. B. Hume, C. W. Chou, T. Rosenband, and D. J. Wineland, Phys. Rev. A \textbf{80}, 052302 (2009).

\bibitem{Vitanov} N. V. Vitanov, B. W. Shore, L. Yatsenko, K. B\"{o}hmer, T. Halfmann, T. Rickes, and K. Bergmann, Opt. Commun. \textbf{199}, 117 (2001).

\end{thebibliography}
\end{document}